\documentclass[pre,12pt]{revtex4}

\usepackage{graphicx}   
\usepackage{verbatim}   
\pdfoutput=1 

\begin{document}

\title{Entropy of Difference}
\author{Pasquale Nardone}
\email{pnardon@ulb.ac.be}
\affiliation{Physics Department, Universit\'e Libre de Bruxelles}
\address{50 av F. D. Roosevelt, 1050 Bruxelles, Belgium}
\date{\today}

\begin{abstract}
Here, we propose a new tool to estimate the complexity of a time series: the entropy of difference (ED). The method is based solely on the sign of the difference between neighboring values in a time series. This makes it possible to describe the signal as efficiently as prior proposed parameters such as permutation entropy (PE) or modified permutation entropy (mPE), but (1) reduces the size of the sample that is necessary to estimate the parameter value, and (2) enables the use of the Kullback-Leibler divergence to estimate the ÒdistanceÓ between the time series data and random signals.
\end{abstract}

\maketitle

\section{Introduction}
Permutation entropy (PE), introduced by  Bandt and Pompe\cite{BB}, as well as its modified version\cite{mPE}, are both efficient tools to measure the complexity of chaotic time series. Both methods propose to analyze time series: $X=\{x_1,x_2,\cdots x_k\cdots\}$ by first  choosing an embedding dimension $m$ to split the original data in a subset of $m$-tuples: $\{\{x_1,x_2\cdots x_{m}\},\{x_2, x_3, \cdots x_{1+m}\},\cdots\}$, then to substitute to the $m$-tuples values by the rank of the values, resulting in a new symbolic representation of the time series. For example, consider the time series $X=\{0.2, 0.1, 0.6, 0.4, 0.1, 0.2, 0.4, 0.8, 0.5, 1., 0.3, 0.1,\cdots\}$. Choosing, for example, an embedding dimension $m=4$, will split the data in a set of $4$-tuples: $X_4=\{\{0.2, 0.1, 0.6, 0.4\}, \{0.1, 0.6, 0.4, 0.1\}, \{0.6, 0.4, 0.1, 0.2\},\cdots\}$. The Bandt-Pompe method will associate the rank of the value with each $4$-tuples. Thus, in $\{0.2, 0.1, 0.6, 0.4\}$ the lowest element $0.1$ is in position $2$, the second element $0.2$ is in position $1$, $0.4$ is in position $4$ and finally $0.6$ is in position $3$. Thus the 4-tuple $\{0.2, 0.1, 0.6, 0.4\}$ is rewritten as $\{2,1,4,3\}$. This procedure thus results in each  $X_4$ to be rewritten as a symbolic list:$\{\{2, 1, 4, 3\}, \{1, 4, 3, 2\}, \{3, 4, 2, 1\}\cdots\}$. Each element is then a permutation $\pi$ of the set $\{1,2,3,4\}$. 
Next, the probability of each permutation $\pi$ in $X_m$ is then computed: $p_m(\pi)$, and finally the PE for the embedding dimension $m$, is defined as $\hbox{PE}_m(X)=-\sum_\pi p_m(\pi)\log(p_m(\pi))$. The modified permutation entropy (mPE) just deals with those cases in which equal quantities may appear in the $m$-tuples. For example for the $m$-tuple $\{0.1, 0.6, 0.4, 0.1\}$, computing PE will produce $ \{1, 4, 3, 2\}$ while computing mPE will associate $\{1, 1, 3, 2\}$\footnote{see appendix 1}. Both methods are widely used due to their conceptual and computational simplicity\cite{lot, lot1, lot2, lot3, lot4, lot5}. For random signals, PE leads to a constant probability $q_m(\pi)=1/m!$, which does not make it possible to evaluate the ``distance" between the probability found in the signal: $p_m(\pi)$ and the probability produced by a random signal: $q_m$, with the Kullback-Leibler (KL) divergence\cite{KL, RE}: ${\tt KL}_m(p\| q)=\sum_\pi p_m(\pi)\log_2(p_m(\pi)/q_m(\pi))$. Furthermore, the number ${\tt K}_m$ of  $m$-tuples are $m!$ for PE and even greater for mPE\cite{mPE}, thus requiring then a large data sample to perform  significant statistical estimation of $p_m$.

\section{Entropy of difference-method}
The entropy of difference (ED) method proposes  to substitute to the $m$-tuples with strings $s$ containing the sign (``+" or ``-"), representing of the difference between subsequent elements in the $m$-tuples. For the same $X_4$: $\{\{0.2, 0.1, 0.6, 0.4\}, \{0.1, 0.6, 0.4, 0.1\}, \{0.6, 0.4, 0.1, 0.2\},\cdots\}$ this leads to the representation :  $\{``-+-", ``+--",``--+",\cdots\}$. For an $m$ value, we have $2^{m-1}$ strings from $``+++\cdots +"$ to $``---\cdots-"$. Again we compute, in the time series,  the probability distribution $p_m(s)$ of these strings $s$ and define the entropy of difference of order $m$ as : $\hbox{ED}_m=-\sum_s p_m(s)\log p_m(s)$. The number of elements: ${\tt K}_m$  to be treated, for an embedding $m$, are smaller for ED compared with the number of permutations $\pi$ in PE or to the elements in mPE (see table I).

\begin{table}[ht]
\caption{\label{tab:mPE} {\tt K} values, for different $m$-embedding}
\begin{ruledtabular}
\begin{tabular}{cccccc}
$m$&$3$&$4$&$5$&$6$&$7$\\
\hline
${\tt K}_{PE}$& 6&24&120 & 720& 5040\\
${\tt K}_{mPE}$& 13&73&501&4051&37633\\
${\tt K}_{ED}$& 4&8&16&32&64\\
\end{tabular}
\end{ruledtabular}
\end{table}

Furthermore the probability distribution for a string $s$, in a  random signal  : $q_m(s)$ is not constant and could be computed through the recursive equation\footnote{see appendix 2} (in the following equations {\tt x} and {\tt y} are strings):
\begin{eqnarray}
q(+)=q(-)=\frac{1}{2}\nonumber\\
q(\underbrace{+,+,+,\cdots,+}_{m})=\frac{1}{(m+1)!}\nonumber\\
q(-,\hbox{\tt x})=q(\hbox{\tt x})-q(+,\hbox{\tt x})\nonumber\\
q(\hbox{\tt x},-)=q(\hbox{\tt x})-q(\hbox{\tt x},+)\nonumber\\
q(\hbox{\tt x}, -,\hbox{\tt y})=q(\hbox{\tt x})q(\hbox{\tt y})-q(\hbox{\tt x},+,\hbox{\tt y})\nonumber\\
\end{eqnarray}
leading to a complex probability distribution. For example for $m=9$ we have $2^8= 256$ strings with the highest probability for the $``+-+-+-+-"$ string (and its symmetric  $``-+-+-+-+"$): $q_9(\hbox{\tt max})=\frac{62}{2835}\approx 0.02187$ (see Fig. I). These probabilities $q_m(s)$ could then be used  to determine the KL-divergence between the time series probability $p_m(s)$ and the random signal. 

\begin{figure}[ht]
\includegraphics{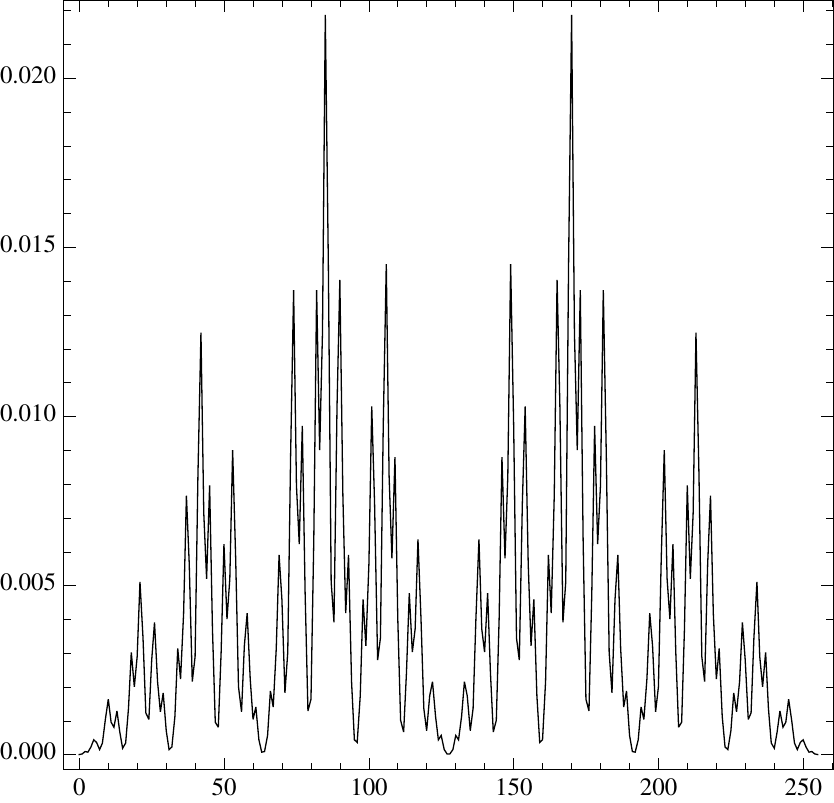}
\caption{\label{fig:prob} The $2^8$ values for the probability of $q_9(s)$, from $s=---...\equiv0$ to $s=+++...\equiv255$}
\end{figure}

Despite the complexity of $q_m(s)$, the Shannon entropy for a random signal : $-\sum_s q_m(s)\log_2 q_m(s)$ increases linearly with $m$, with a slope $\approx0.905$.

\begin{figure}[ht]
\includegraphics{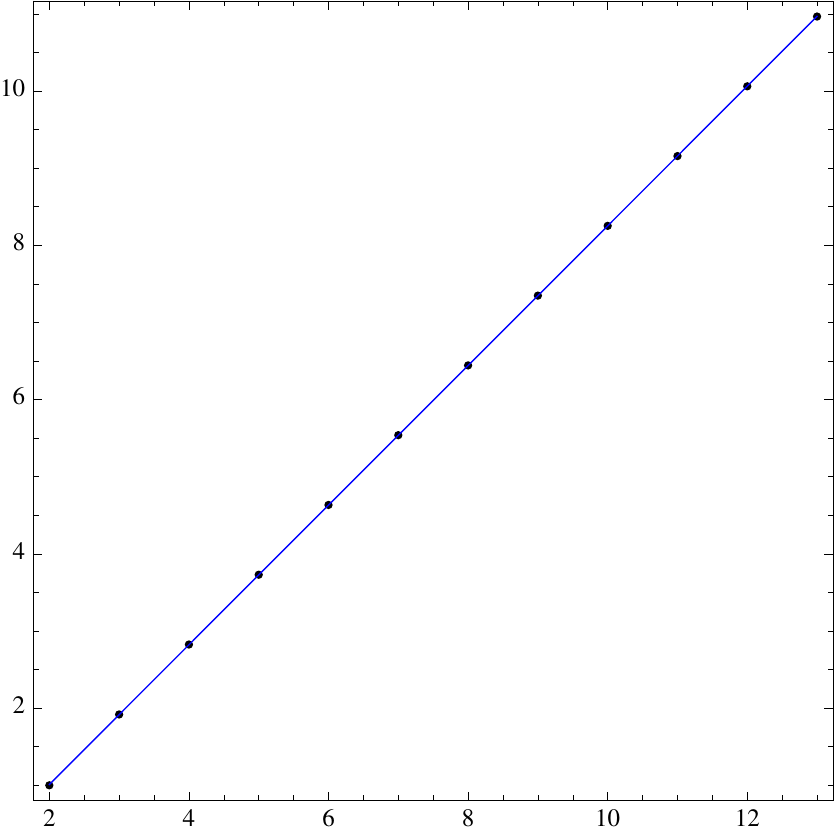}
\caption{\label{fig:fit} The Shannon entropy of $q_m(s)$ increases linearly with $m$, the fit $-0.799574 + 0.905206\;m$ gives a   sum of squared residuals of $1.7\;10^{-4}$ and a p-value=$1.57\;10^{-12}$ and $1.62\;10^{-30}$ on the fit parameter respectively.}
\end{figure}

\section{Chaotic logistic map example}
Let us illustrate the use of ED on the well know logistic map\cite{lo} ${\tt Lo}(x,\lambda)$ driven by the parameter $\lambda$. 
\begin{equation}
x_{n+1}={\tt Lo}(x_n,\lambda)=\lambda x_n (1-x_n)
\end{equation}
It is obvious that for a range of values of $\lambda$ where the time series reaches a periodic behavior (any cyclic oscillation between $n$ different values), the ED will remain constant. The evaluation of the ED could thus be used as a new complexity parameter to determine the behavior of the time series (see FIG. 3).

\begin{figure}[ht]
\includegraphics[width=480pt]{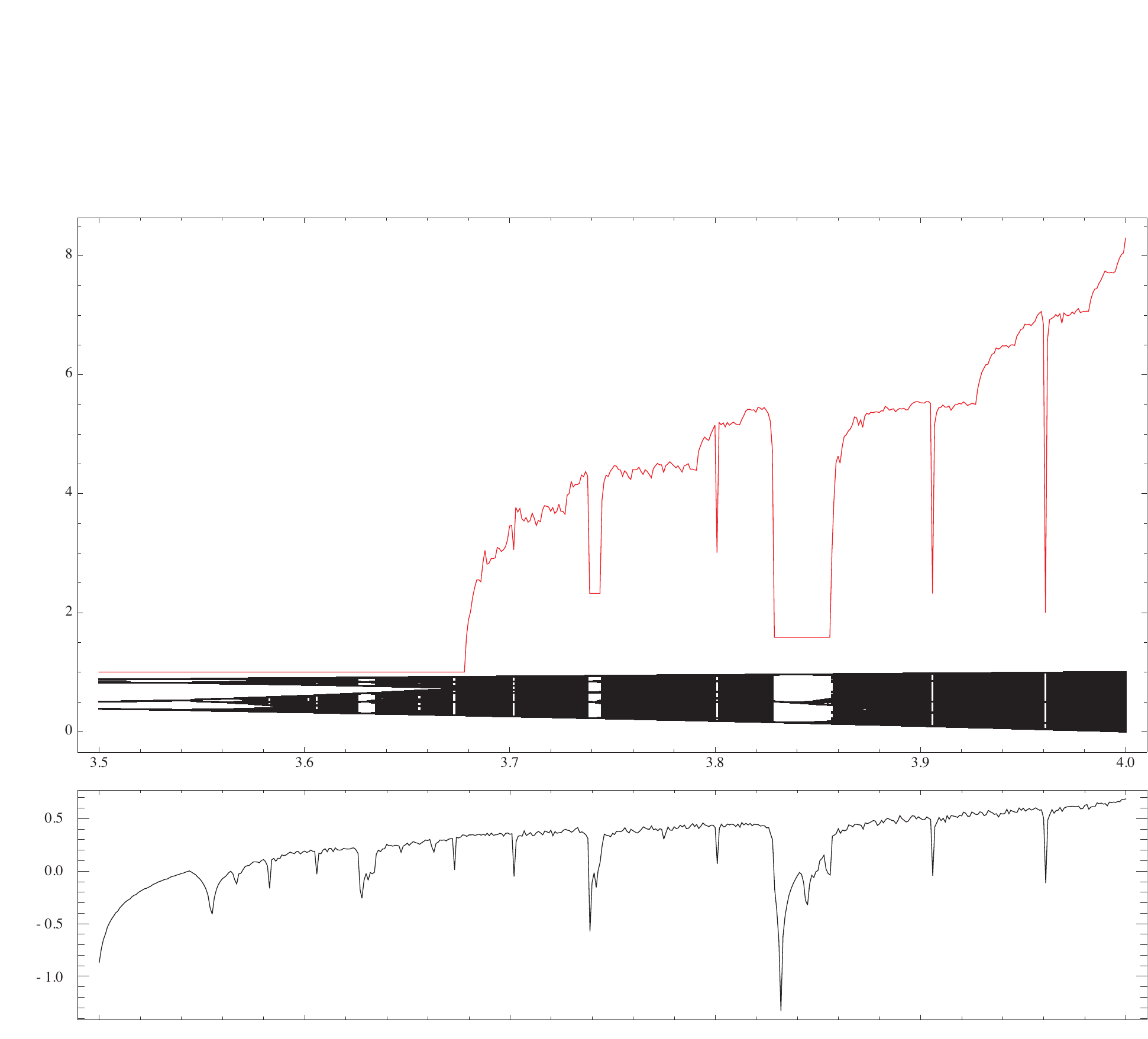}
\caption{\label{fig:Bif} The $\hbox{\tt ED}_{13}$ (strings of length 12) is plotted versus $\lambda$, with the bifurcation diagram, and the value of the Lyapunov exponent respectively. The constant value appears when the logistic map enter into a periodic regime.
}
\end{figure}

For $\lambda=4$ we know that the data are randomly distributed with a probability density given by\cite{log}
\begin{equation}
p_{{\tt Lo}}(x)=\frac{1}{\pi  \sqrt{(1-x) x}}
\end{equation}
We can then compute exactly the {\tt ED} for an $m$-embedding, and the {\tt KL}-divergence from a random signal.
For example, for $m=2$, we can determine the $p_+$ and $p_-$  by solving the inequality $x<{\tt Lo}(x)$ and $x>{\tt Lo}(x)$ respectively which implies that $0<x<3/4$ and $3/4<x<1$, and then
\begin{equation}
p_+=\int_0^{3/4} dx\ p_{{\tt Lo}}(x)=\frac{2}{3}\quad p_-=\int_{3/4}^{1} dx\ p_{{\tt Lo}}(x)=\frac{1}{3}
\end{equation}
In this case the logistic map produces a signal that contains twice as many increasing pairs $``+"$ than decreasing pairs $``-"$. 
So:
\begin{equation}
\hbox{\tt ED}_2=-(\frac{2}{3}\log_2\frac{2}{3}+\frac{1}{3}\log_2\frac{1}{3})=\log_2\frac{3}{2^{2/3}} \approx 0.918\quad
\hbox{\tt KL}_2=\frac{1}{3}\log_2\frac{32}{27}\approx 0.082
\end{equation}
For $m=3$ and $m=4$ we can perform the same calculation:
\begin{eqnarray}
p_3(++)=\frac{1}{3}\quad p_3(+-)=\frac{1}{3}\quad p_3(-+)=\frac{1}{3}\\ \to \hbox{\tt ED}_3=\log_2 3\approx 1.58
\quad\hbox{KL}_3=\frac{1}{3}\approx 0.33\nonumber
\end{eqnarray}
Effectively the logistic map with $\lambda=4$ forbids the string ``- -" where $x_1>x_2>x_3$. For strings of length $3$ we also have also the non zero values:
\begin{eqnarray}
p_4(+++)=p_4(++-)=p_4(-++)=p_4(-+-)=\frac{1}{6}\quad p_4(+-+)=\frac{2}{6}
\nonumber\\ \to\hbox{\tt ED}_4=\log_2 108^\frac{1}{3}\approx 2.25\quad\hbox{\tt KL}_4=\log_2\left({16384\over 1125}\right)^{1/6}\approx 0.64
\end{eqnarray}
The probability of difference $p_m(s)$ for some string length $m$ versus $s$ the string binary value, where ``+"$\rightarrow 1$ and ``-"$\rightarrow 0$, give us the ``spectrum of difference" for the distribution $p$ (see FIG. 4). 

\begin{figure}[ht]
\includegraphics[scale=0.8]{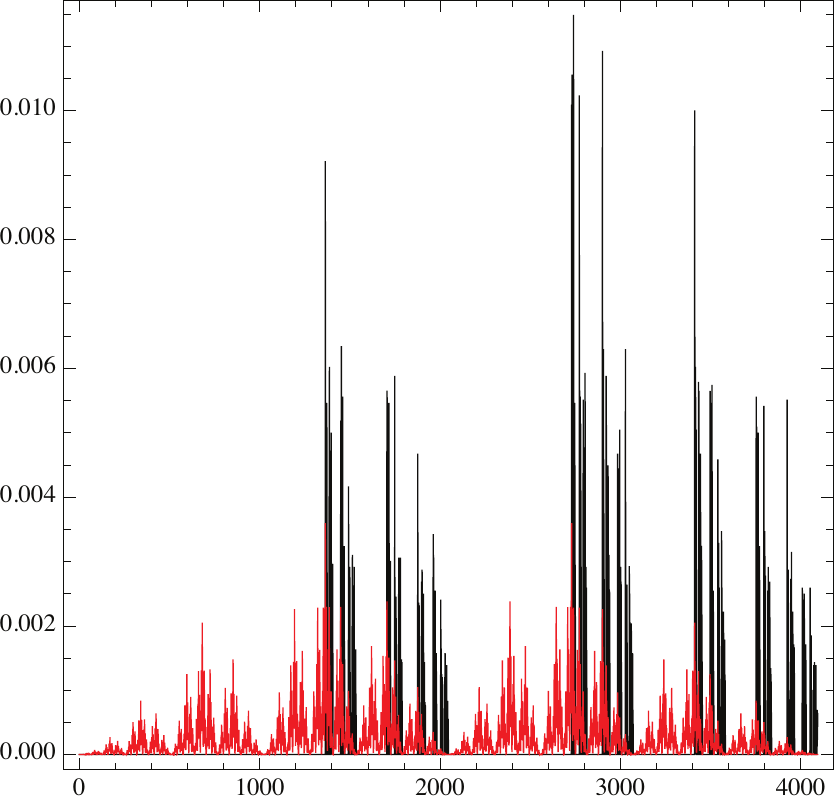}
\caption{\label{fig:spectrumLogistic} The spectrum of $p_{13}$ versus the string binary value (from $0$ to $2^{12}-1$) for the logistic map at $\lambda=4$ and the one from a random distribution $q_{13}$ }
\end{figure}

\section{$\hbox{\tt KL}_m(p|q)$ divergences versus $m$ on real data and on maps}

The manner in which the $\hbox{\tt KL}_m(p|q)$ evolves with $m$ is another parameter of the complexity measure. $\hbox{\tt KL}_m(p|q)$  measures the loss of informations when the random distribution $q_m$ is used to predict the distribution $p_m$. Increasing $m$ introduces more bits information in the signal and the behavior versus $m$ shows how the data diverges from a random distribution.

The graphics (see FIG. 5) shows the behavior of $\hbox{\tt KL}_m$ versus $m$ for two different chaotic maps and for real financial data\cite{finance} : the opening value of the {\tt nasdaq100}, {\tt bel20}  everyday from 2000 to 2013. 
For maps, the logarithmic map $x_{n+1}=\ln(a|x_n|)$ and logistic map are shown. 

For maps the simulation starts with a random number between 0 and 1, then first iterate 500 times to avoid transients. Starting with that seeds, 720 iterates where kept on which the $\hbox{\tt KL}_m$ where computed. It can be seen that the Kullback-Leibler divergence from the logistic map at $\lambda=4$ to the random signal is fitted by a quadratic function of $m$: $\hbox{\tt KL}_m=-0.4260+0.2326\;m+0.0095\; m^2$ (p-value$\approx 2\;10^{-7}$ for all the parameter), while the logarithmic map behavior is linear in the range $a\in[0.4,2.2]$. Financial data are also quadratic $\hbox{\tt KL}_m(\hbox{\tt nasdaq})=0.1824-0.0973\;m+0.0178\; m^2$, $\hbox{\tt KL}_m(\hbox{\tt bel20})=0.1587-0.0886\;m+0.0182\; m^2$ with a higher curvature than the logistic map due to the fact that the spectrum of the probability $p_m$ is compatible with a constant distribution (see FIG. 6) rendering the prediction of increase or decrease signal completely random, which is not the case in any true random signal.

\begin{figure}[h!]
\includegraphics[scale=1]{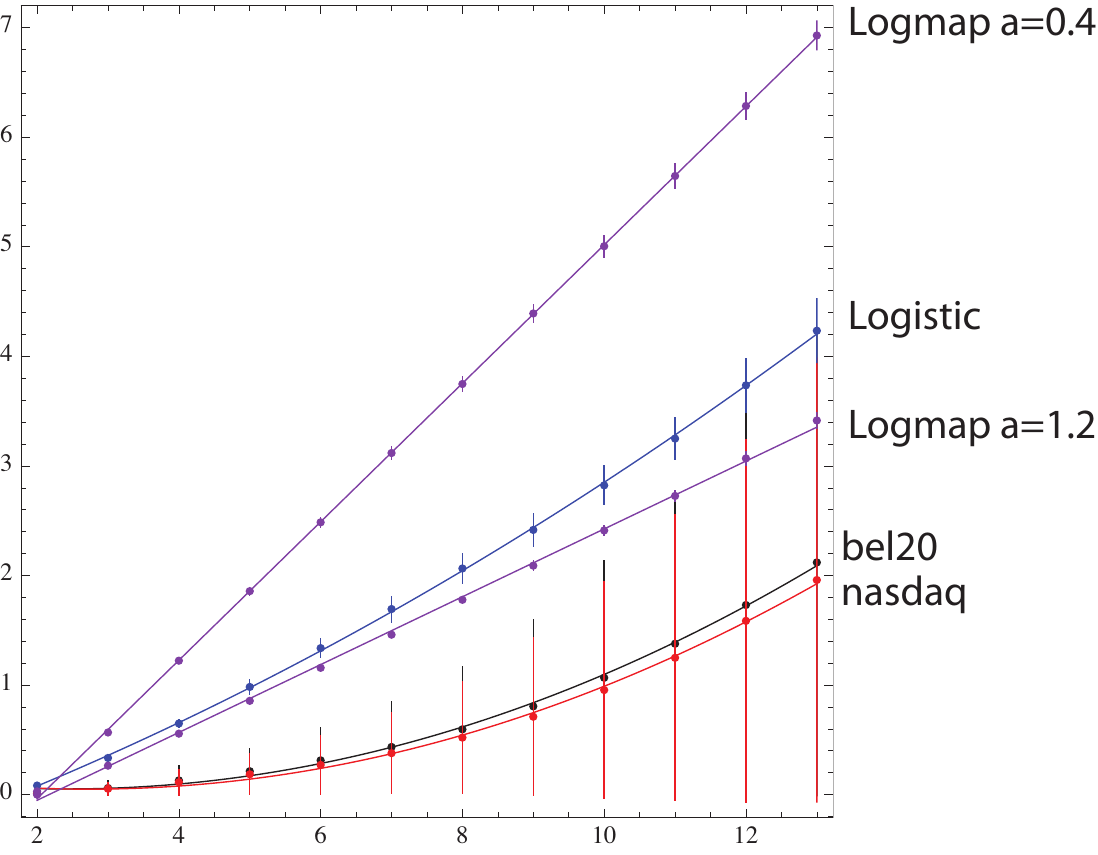}
\caption{\label{fig:figKL} The KL-divergence for the data}
\end{figure}

\begin{figure}[h!]
\includegraphics[scale=1]{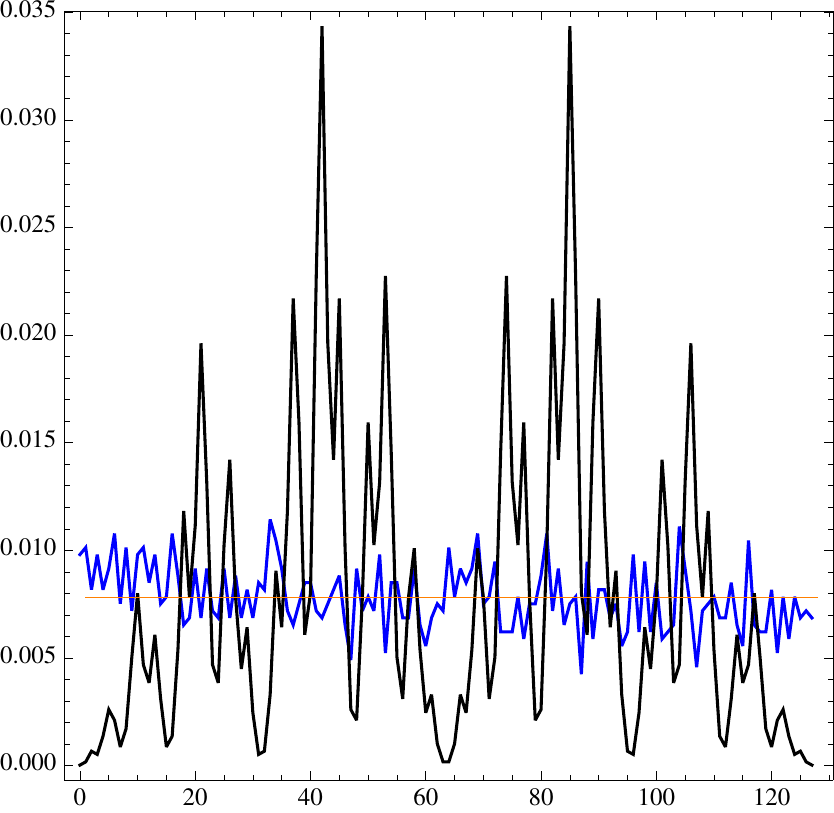}
\caption{\label{fig:BEL20} The spectrum of $p_8$ versus the string binary value (from $0$ to $2^7-1$) for the bel20 financial data}
\end{figure}

\section{Conclusions}
The simple property of increases or decreases in a signal makes it possible to introduce the entropy of difference $\hbox{\tt ED}_m$ as a new efficient complexity measure for chaotic time series. The probability distribution of string $q_m$ for random signal is used to evaluate the Kullback-Leibler divergence versus the number of data $m$ used to build the difference string. This $\hbox{\tt KL}_m$ shows different behavior for different types of signal and can also be used also to characterize the complexity of a time series.

\appendix*
\section{1}
The Mathematica program for $m$-embding, PE and mPE are  simple:
\begin{verbatim}
mEmbedding[Xlist_,m_]:=Partition[Xlist,m,1];
PE[mList_]:=Ordering[mList];
mPE[mList_]:=Flatten[Map[First[Position[mList, #]] &, Sort[mList]]];
\end{verbatim}

\appendix*
\section{2}
The Mathematica program for the probability $q(s)$:
\begin{verbatim}
P["+"]= P["-"] = 1/2;
P["-", x__] := P[x] - P["+", x];
P[x__, "-"] := P[x] - P[x, "+"];
P[x__, "-", y__] := P[x] P[y] - P[x, "+", y];
P[x__] :=1/(StringLength[StringJoin[x]] + 1)!
\end{verbatim}

\bibliographystyle{plain}	
\bibliography{EntropyOfDifference}		
\end{document}